# Estimating Participation Factors and Mode Shapes for Electromechanical Oscillations in Ambient Conditions

Xiaozhe Wang, *Member, IEEE,* Ilias Zenelis

*Abstract*—In this paper, a new technique is applied to conduct mode identification using ambient measurement data. The proposed hybrid measurement- and model-based method can accurately estimate the system state matrix in ambient conditions, the eigenvalues and eigenvectors of which readily provide all the modal knowledge including frequencies, damping ratios, mode shapes, and more importantly, participation factors. Numerical simulations show that the proposed technique is able to provide accurate estimation of modal knowledge for all modes. In addition, the discrepancy between the participation factor and the mode shape is shown through a numerical example, demonstrating that using the mode shape may not effectively pinpoint the best location for damping control. Therefore, the proposed technique capable of estimating participation factors may greatly facilitate designing damping controls.

## I. INTRODUCTION

Accurate estimation of low-frequency electromechanical oscillations is of paramount significance to ensure the secure operation of any power system. There are typically two approaches to conduct such estimation: linearizing the complete power system dynamic model around its steady-state operating point or utilizing measurement data to extract modal information (e.g., [1], [2]). Since an accurate model along with parameter values is usually unavailable in practice, much interests have been given to develop measurement-based methods utilizing high-sampled and synchronized data from phasor measurement units (PMUs). PMU data can be further categorized into ring-down data and ambient data. Particularly, in this paper, the proposed technique utilizes ambient PMU data that can be easily obtained in normal operating condition without any large disturbance.

In previous study, a large number of mode shape estimation methods using ambient data have been developed, including the spectral method [3], the frequency domain decomposition (FDD) method [4], the channel matching method [5], the transfer function (TF) method [6], and the subspace methods [7], [8]. Particularly, it has been shown in [9] that the subspace methods may achieve more accurate estimations in high-damping situation and in the presence of oscillatory disturbances compared with the other methods, and hence have received wide interests and applications. However, there are some limitations of the subspace methods. First, the estimated state matrix by subspace methods is not the true system state matrix; therefore, the left eigenvector associated with the mode of interest can not be estimated, leading to unattainability of the participation factor [10]. Second, the subspace methods are computationally expensive because they require singular value decomposition (SVD) of large-dimensional matrices.

In this paper, we will apply a novel method proposed in [11] [12] to carry out mode identifications for electromechanical oscillations, which may potentially overcome the challenges faced by the subspace methods. The proposed method is able to accurately estimate the true system state matrix and thus all the modal knowledge including frequencies, damping ratios, mode shapes, and more importantly, **participation factors** that carry crucial information for control. To the knowledge of the authors, there have been no measurement-based methods on estimating participation factors without knowing the detailed network model. In addition, the proposed mode identification method is computationally efficient and can be implemented in near real-time.

Simulation results demonstrate that the proposed mode identification method can accurately estimate the modal knowledge (e.g., frequencies, damping ratios, mode shapes, participation factors) for **all modes**. Despite of the fact that the method is developed based on the classical models, it shows good applicability to high-order generator models with detailed control. Furthermore, the proposed method is robust against measurement noise and thus is applicable in real-world applications. The last but not the least, we show a potential discrepancy between the mode shape and the participation factor regarding the components with largest magnitudes, which implies that relying on the mode shape may not lead to the best locations for damping control. Therefore, the proposed technique capable of estimating participation factors may greatly facilitate damping control design by pinpointing the most problematic components, i.e., the best locations.

The rest of the paper is organized as follows. Section II introduces the power system stochastic dynamic model and elaborates the proposed mode identification method. Section III presents the results of numerical studies.

## II. THE METHODOLOGY FOR MODE IDENTIFICATION

In this paper, we are interested in the case when the power system operates in ambient condition and its dynamics can be

This work is supported by Natural Sciences and Engineering Research Council (NSERC) Discovery Grant, NSERC RGPIN-2016-04570.

Xiaozhe Wang and Ilias Zenelis are with the Department of Electrical and Computer Engineering, McGill University, Montréal, QC H3A 0G4, Canada. email: xiaozhe.wang2@mcgill.ca, ilias.zenelis@mail.mcgill.ca

described by the classical generator model:

$$\dot{\boldsymbol{\delta}} = \boldsymbol{\omega} \qquad (1)$$

$$M\dot{\boldsymbol{\omega}} = \boldsymbol{P_m} - \boldsymbol{P_e} - D\boldsymbol{\omega} \qquad (2)$$

with

$$P_{e_i} = \sum_{j=1}^{n} E_i E_j (G_{ij}\cos(\delta_i - \delta_j) + B_{ij}\sin(\delta_i - \delta_j)) \qquad (3)$$

where $\boldsymbol{\delta}$, $\boldsymbol{\omega}$ are generator rotor angles and relative rotor speeds w.r.t. the synchronous speed; $M = \text{diag}(M_1, ...M_n)$, $D = \text{diag}(D_1, ...D_n)$ are inertia and damping constants, respectively; $\boldsymbol{P_m}$ and $\boldsymbol{P_e}$ are generators' mechanical power and electrical power; $E = \text{diag}([E_1, ..., E_n])$ are the magnitudes of electromotive force behind the transient reactance; $G_{ij} + jB_{ij}$ is the $(i,j)$th entry of the reduced admittance matrix that includes generators' impedances.

Similar to the approach in [13], we assume that the system loads modeled as constant impedances are experiencing Gaussian variations which can be translated to the diagonal elements of the reduced admittance matrix as below:

$$Y(i,i) = Y_{ii}(1 + \sigma_i dW_i)\angle\phi_{ii} \qquad (4)$$

where $W$ is a standard Wiener process and $\sigma_i^2$ is the variance of load variations. Therefore, the system model takes the following form [11] [13]:

$$\dot{\boldsymbol{\delta}} = \boldsymbol{\omega} \qquad (5)$$

$$M\dot{\boldsymbol{\omega}} = \boldsymbol{P_m} - \boldsymbol{P_e} - D\boldsymbol{\omega} - E^2 G\Sigma\boldsymbol{\xi} \qquad (6)$$

where $\boldsymbol{\xi} = \dot{W}$ is a vector of independent standard Gaussian random variables, and $\Sigma = \text{diag}([\sigma_1, ..., \sigma_n])$ denotes the variation intensity. Note that the term $E^2 G\Sigma\boldsymbol{\xi}$ translates the stochastic variation at load side to the stochastic variation of electrical power at generator side [13].

Representing (5)-(6) in a compact form as a set of stochastic differential equations:

$$\dot{\boldsymbol{x}} = A\boldsymbol{x} + B\boldsymbol{\xi} \qquad (7)$$

where

$$\boldsymbol{x} = [\boldsymbol{\delta}, \boldsymbol{\omega}]^T \qquad (8)$$

$$A = \begin{bmatrix} 0 & I_n \\ -M^{-1}\frac{\partial \boldsymbol{P_e}}{\partial \boldsymbol{\delta}} & -M^{-1}D \end{bmatrix} \qquad (9)$$

$$B = [0, -M^{-1}E^2 G\Sigma]^T \qquad (10)$$

Specifically, $\boldsymbol{x}$ is a multivariate Ornstein-Uhlenbeck process [14]. For a generic Ornstein-Uhlenbeck process, the stationary covariance matrix $C_{xx} = \begin{bmatrix} C_{\delta\delta} & C_{\delta\omega} \\ C_{\omega\delta} & C_{\omega\omega} \end{bmatrix}$ satisfies the following Lyapunov equation if the system state matrix $A$ is stable (typically satisfied in ambient conditions) [14] [15]:

$$AC_{xx} + C_{xx}A^T = -BB^T \qquad (11)$$

This relationship combines the statistical properties of states that can be estimated from PMU data and the physical model knowledge, providing an ingenious way to estimate model information from measurements. Substituting the detailed expressions of $A$ and $B$ to (11), we obtain:

$$(\frac{\partial \boldsymbol{P_e}}{\partial \boldsymbol{\delta}}) = MC_{\omega\omega}C_{\delta\delta}^{-1} - DC_{\omega\delta}C_{\delta\delta}^{-1} \qquad (12)$$

This derived relation indicates that the dynamic state jacobian matrix $\frac{\partial \boldsymbol{P_e}}{\partial \boldsymbol{\delta}}$ can be estimated from the statistics of state variables extracted from the PMU measurements given that the damping factors $D$ and inertias $M$ of generators are known. Furthermore, we can readily construct the system state matrix $A$ by (9) from which the electromechanical modes, mode shapes and participation factors can be obtained.

### A. The estimation of the covariance matrices

The stationary covariance matrix is defined as:

$$C_{\delta\delta} = \begin{bmatrix} C_{\delta_1\delta_1} & C_{\delta_1\delta_2} & \dots & C_{\delta_1\delta_n} \\ C_{\delta_2\delta_1} & C_{\delta_2\delta_2} & \dots & C_{\delta_2\delta_n} \\ \vdots & \vdots & \ddots & \vdots \\ C_{\delta_n\delta_1} & C_{\delta_n\delta_2} & \dots & C_{\delta_n\delta_n} \end{bmatrix} \qquad (13)$$

where $C_{\delta_i\delta_j} = \text{E}[(\delta_i - \mu_i)(\delta_j - \mu_j)]$, and $\mu_i$ is the mean of $\delta_i$. However, $C_{\delta\delta}$ is typically unknown in practice due to limited data. Therefore, sample covariance matrix $Q_{\delta\delta}$ is used to estimate $C_{\delta\delta}$, each entry of which is calculated as below [14]:

$$Q_{\delta_i\delta_j} = \frac{1}{N-1}\sum_{k=1}^{N}(\delta_{ki} - \bar{\bar{\delta}}_i)(\delta_{kj} - \bar{\bar{\delta}}_j) \qquad (14)$$

where $\bar{\bar{\delta}}_i$ denotes the sample mean of $\delta_i$, and $N$ is the sample size. Likewise, $C_{\omega\omega}$, $C_{\omega\delta}$ can be estimated by $Q_{\omega\omega}$, $Q_{\omega\delta}$. A window size around 500s is used in the examples of this paper, which shows good accuracy. Like other measurement-based methods, there is always a tradeoff between estimation accuracy and sample size.

As a result, the dynamic Jacobian matrix can be obtained from:

$$\left(\frac{\partial \boldsymbol{P_e}}{\partial \boldsymbol{\delta}}\right) = MQ_{\omega\omega}Q_{\delta\delta}^{-1} - DQ_{\omega\delta}Q_{\delta\delta}^{-1} \qquad (15)$$

if generator inertias $M$ and damping constants $D$ are known.

### B. The mode identification algorithm

We assume that PMUs are deployed at all the substations that generators are connected to. This assumption may be too optimistic currently, while is fairly reasonable in the near future considering the increasing rate of deployment of PMUs. Rotor angle $\boldsymbol{\delta}$ and rotor speed $\boldsymbol{\omega}$ in steady state can be calculated from PMU measurements. Discussion how exactly it is done is beyond the scope of this paper and the readers are directed to, e.g., [16]. If the inertias $M$ and damping $D$ of generators are known, the following algorithm identifies and estimates electromechanical oscillations via ambient PMU measurements:

**Step 1.** Compute $\boldsymbol{\delta}$ and $\boldsymbol{\omega}$ from the PMUs, and estimate the sample covariance matrices $Q_{\delta\delta}$, $Q_{\omega\omega}$ and $Q_{\omega\delta}$.

**Step 2.** Estimate the Jacobian matrix $\frac{\partial \boldsymbol{P_e}}{\partial \boldsymbol{\delta}}$ by $(\frac{\partial \boldsymbol{P_e}}{\partial \boldsymbol{\delta}}) = MQ_{\omega\omega}Q_{\delta\delta}^{-1} - DQ_{\omega\delta}Q_{\delta\delta}^{-1}$, and construct the system state matrix $A = \begin{bmatrix} 0 & I_n \\ -M^{-1}\frac{\partial \boldsymbol{P_e}}{\partial \boldsymbol{\delta}} & -M^{-1}D \end{bmatrix}$.

**Step 3.** Compute the eigenvalues $\Lambda$, the right eigenvectors $\Phi = [\phi_1, \ldots, \phi_{2n}]$ and the left eigenvectors $\Psi = [\psi_1^T, \ldots \psi_{2n}^T]^T$ of $A$. The mode frequencies and the damping ratios can be directly obtained from $\Lambda$; mode shapes correspond to $\Phi$; the participation factor of the mode $\lambda_i$ can be calculated by $P_i = [P_{1i}, \ldots P_{2n,i}]^T = [\phi_{1i}\psi_{i1}, \ldots, \phi_{2n,i}\psi_{i,2n}]^T$.

## III. RESULTS AND ANALYSIS

Two cases in the IEEE 39-bus 10-generator system are to be presented. The first case is to validate the proposed mode identification method using the second-order generator models. The second case is to illustrate that the proposed method works well for high-order generator models with detailed control strategies, even though it is developed based on the classical generator models. In addition, it is shown that the method is robust against measurement noise, indicating good feasibility in practical applications.

The parameters of the two cases can be found in: https://github.com/zenili/Mode-Participation-Estimation-2017. All simulations were done in PSAT-2.1.9 [17].

### A. Case I: Validation on Classical Generator Model

In the 39-bus system, 10 generators are modelled as the second-order models as in (1)-(2); the angle of Generator 1 serves as the reference. We set $\sigma_i = 5$ for all generators describing the stochastic load variations in the system. The sampling rate is 20 samples comparable with typical PMU sampling rate. The emulated trajectories of the rotor angle and the rotor speed of Generator 2 obtained from the PMU data are presented in Fig. 1, from which we see that system is stable and the state variables are fluctuating around their nominal operating states.

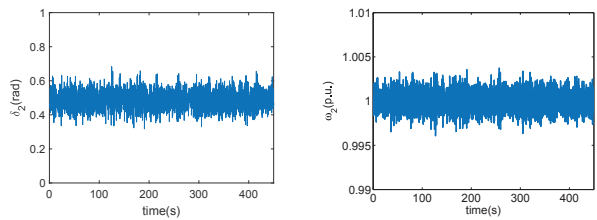

Fig. 1: Trajectories of $\delta_2$ and $\omega_2$ in the 39-bus system using the classical models.

Applying the proposed mode identification algorithm using the 450s sample data shown in Fig. 1, we obtain the estimation results of mode frequencies and damping ratios as presented in Table I-II. It is observed that the proposed method is able to correctly distinguish different modes and accurately estimate their frequencies and damping ratios, even though the frequencies of some modes are close to each other (e.g., Mode 2-4) and some damping ratios are high (e.g., Mode 1). To further illustrate the accuracy of the proposed technique, we present a comparison between the estimated eigenvalues and the true eigenvalues of the system state matrix $A$ in Fig. 2 (only the positive frequencies are shown), from which we see that all eigenvalues can be well estimated with relatively good accuracy.

| Mode | True frequency (Hz) | Estimated frequency (Hz) | Estimation error (%) |
|------|---------------------|--------------------------|----------------------|
| 1 | 0.551 | 0.553 | 0.30 |
| 2 | 2.574 | 2.573 | 0.05 |
| 3 | 2.505 | 2.494 | 0.45 |
| 4 | 2.449 | 2.425 | 0.97 |
| 5 | 1.337 | 1.336 | 0.06 |
| 6 | 1.489 | 1.482 | 0.5 |
| 7 | 1.671 | 1.663 | 0.54 |
| 8 | 1.891 | 1.867 | 1.27 |
| 9 | 1.828 | 1.829 | 0.06 |

TABLE I: Case I: the estimation results of frequencies for all modes.

| Mode | True damping ratio(%) | Estimated damping ratio (%) | Estimation error (%) |
|------|------------------------|-----------------------------|----------------------|
| 1 | 6.12 | 6.03 | 1.40 |
| 2 | 0.20 | 0.19 | 4.55 |
| 3 | 1.14 | 1.15 | 0.89 |
| 4 | 2.44 | 2.48 | 1.37 |
| 5 | 2.07 | 2.16 | 4.39 |
| 6 | 2.34 | 2.36 | 0.96 |
| 7 | 1.09 | 1.03 | 5.56 |
| 8 | 2.58 | 2.62 | 1.77 |
| 9 | 2.99 | 2.97 | 0.49 |

TABLE II: Case I: the estimation results of damping ratios for all modes.

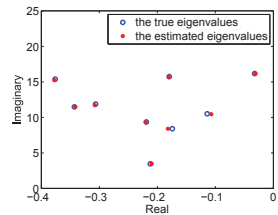

Fig. 2: Case I: a comparison between the estimated eigenvalues and the true eigenvalues.

Regarding mode shapes and participation factors, Fig. 3-4 present the results of the estimated mode shape and the participation factor for Mode 7 which is an inter-area oscillation mode. It can be seen that the estimated mode shape accurately identifies the two groups of oscillation—Generator 4 and 5 are oscillating against Generator 6 and 7. Moreover, the estimated participation factor correctly captures the most influential participants—Generator 5 and 7, implying that the corrective control measure should be adopted at these two generators to depress this particular mode.

### B. Case II: Validation on High-Order Model with Detailed Control and with Measurement Noise

Since the method is developed based on the classical generator model, it is necessary to justify its applicability to higher-order model with detailed control components in real-world power systems. In addition, the PMU data is inevitably subject to measurement noise in real life, the impact of which needs to be investigated.

We still consider the IEEE 39-bus system while all generators are modelled as the third-order models equipped with exciters and power system stabilizers (PSSs). In addition, Gaussian measurement noises with the standard deviation $10^{-3}$ for angles and $10^{-6}$ for speeds are added to the emulated PMU data according to the IEEE standard [18]. Likewise, $450s$ noisy PMU data is employed by the proposed mode identification algorithm, and the estimation results for the mode frequencies and the damping ratios are shown in Table III-IV. It can be seen that the proposed technique still provides good estimation results for frequencies and damping ratios of all modes despite of the fact that higher-order models are used and measurement noises are incorporated. This observation is further confirmed by Fig. 5 that presents a comparison between the estimated eigenvalues and the true ones (only the positive frequencies are shown).

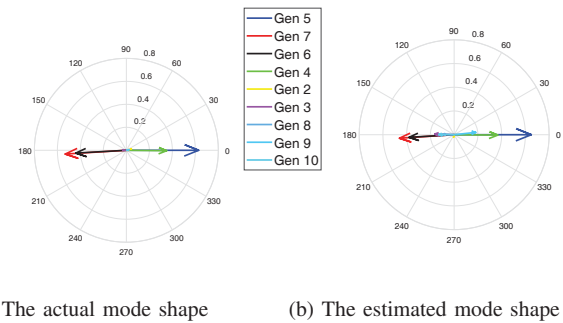

(a) The actual mode shape     (b) The estimated mode shape

Fig. 3: Case I: the actual and estimated mode shapes for Mode 7.

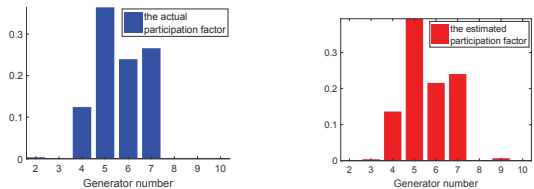

Fig. 4: Case I: the actual and estimated participation factors for Mode 7.

| Mode | True frequency (Hz) | Estimated frequency (Hz) | Estimation error (%) |
|---|---|---|---|
| 1 | 0.497 | 0.498 | 0.29 |
| 2 | 1.267 | 1.270 | 0.22 |
| 3 | 1.425 | 1.413 | 0.81 |
| 4 | 1.500 | 1.501 | 0.08 |
| 5 | 1.650 | 1.642 | 0.51 |
| 6 | 1.817 | 1.797 | 1.06 |
| 7 | 2.182 | 2.157 | 1.15 |
| 8 | 2.052 | 2.050 | 0.097 |
| 9 | 2.110 | 2.082 | 1.33 |

TABLE III: Case II: the estimation results of frequencies for all modes.

Particularly, we consider Mode 2—an inter-area oscillation mode with frequency 1.267 Hz and damping ratio 2.00%. In this mode, Generator 4-7 oscillate against the others. A comparison between the actual and the estimated mode shapes is presented in Fig. 6, showing that the estimated mode shape by the proposed technique well matches with the actual mode shape. In addition, Fig. 7 presents a comparison between the actual participation factor and the estimated one from which we see that the proposed technique accurately estimates the participation factor, indicating that Generator 5 and 9 take the most responsibility in each group for the excitation of this mode.

| Mode | True damping ratio(%) | Estimated damping ratio (%) | Estimation error (%) |
|---|---|---|---|
| 1 | 6.84 | 6.90 | 0.92 |
| 2 | 2.00 | 2.06 | 2.86 |
| 3 | 2.47 | 2.49 | 0.94 |
| 4 | 1.39 | 1.29 | 7.22 |
| 5 | 3.27 | 3.29 | 0.83 |
| 6 | 2.75 | 2.72 | 1.07 |
| 7 | 2.68 | 2.77 | 3.21 |
| 8 | 0.21 | 0.27 | 28.57 |
| 9 | 1.38 | 1.35 | 2.17 |

TABLE IV: Case II: the estimation results of damping ratios for all modes.

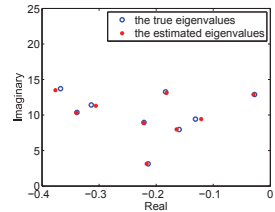

Fig. 5: Case II: a comparison between the estimated eigenvalues and the true eigenvalues.

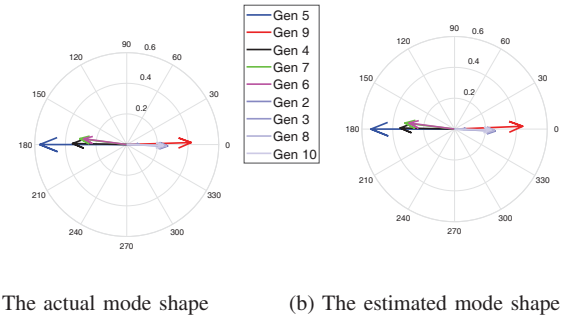

(a) The actual mode shape     (b) The estimated mode shape

Fig. 6: Case II: the actual and estimated mode shapes for Mode 2.

Note that mode shape and participation factor are different in the sense that mode shape gives the relative activity of each state in a particular mode while participation factor presents the influence of each state on a specific mode. Therefore, participation factor may be more effective on designing damping control measures. In most cases and as seen from the previous results, the generators with the largest magnitudes in the mode shape correspond to the generators with the largest magnitudes in the participation factor. That is the reason for which people assume that mode shape can aid in the mitigation of a modal oscillation by pinpointing the right location where a particular mode is most energetic [9]. Nevertheless, we find that the generator with the largest magnitude in the participation factor may not align with the one with the largest magnitude in the mode shape. This discrepancy indicates that using the mode shape may not effectively find the best locations for damping control measures.

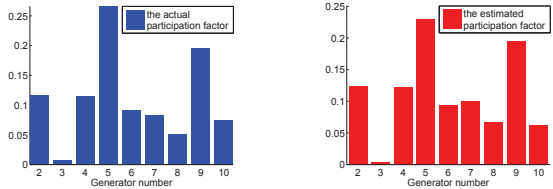

Fig. 7: Case II: the actual and estimated participation factors for Mode 2.

For instance, Mode 6 in the considered 39-bus system is an inter-area mode in which Generator 8 and 10 oscillate against Generator 2 and 9. Fig. 8-9 present the comparisons between its actual mode shape/participation factor and the estimated ones by the proposed technique. It is observed from Fig. 8a and Fig. 9a that Generator 8 has the largest magnitude in the mode shape, yet its component in the participation factor is only the third largest (Generator 2 and 10 are top largest). This difference entails that Generator 2 and 10 have more influence than Generator 8 to Mode 6, even though state variables of Generator 8 are the mostly excited ones in this mode. Therefore, damping control for Mode 6 may be more effectively conducted at Generator 2 and 10 than at Generator 8. Using the mode shape rather than the participation factor may not effectively pinpoint the best location to conduct control measures. From Fig. 9b, it is observed that the proposed algorithm accurately captures the most influential participants—Generator 2 and Generator 10 in the mode, although the influence of Generator 10 is slightly overestimated.

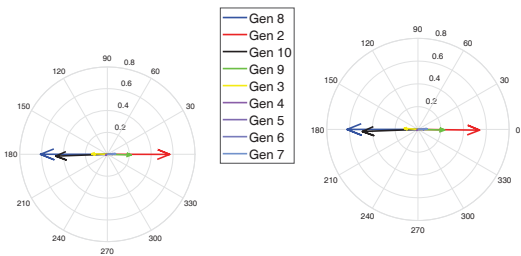

(a) The actual mode shape  (b) The estimated mode shape

Fig. 8: Case II: the actual and estimated mode shapes for Mode 6.

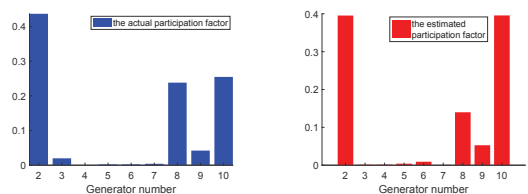

(a) The true participation factor (b) The estimated participation factor

Fig. 9: Case II: the actual and estimated participation factors for Mode 6.

## IV. CONCLUSIONS AND PERSPECTIVES

In this paper, we have proposed a novel mode identification method which can accurately estimate the modal knowledge (e.g., frequencies, damping ratios, mode shapes, participation factors) of electromechanical oscillations using ambient PMU data. Particularly, participation factors can be estimated with reasonably good accuracy without the knowledge of detailed network model and parameters. The potential discrepancy between the participation factor and the mode shape is discussed, showing that the proposed technique capable of estimating participation factor may greatly facilitate effective damping control design. Further efforts are expected to develop damping control measures utilizing the estimated participation factors.